\documentclass[%
 aip,
 jmp,%
 amsmath,
 amssymb,
reprint,%
]{revtex4-1}

\usepackage{graphicx}
\usepackage{dcolumn}
\usepackage{bm}
\usepackage[mathlines]{lineno}
\usepackage{color}%
\usepackage{caption}%
\usepackage{url}

\begin{document}
\title{A low phase noise microwave frequency synthesizer based on parameters optimized NLTL for Cs fountain clocks}%
\author{Wenbing Li}%
\author{Yuanbo Du}%
\thanks{Authors to whom correspondence should be addressed. Electronic addresses: yuanbodu@hust.edu.cn}%
\author{Hui Li}%
\author{Zehuang Lu}%
\affiliation{MOE Key Laboratory of Fundamental Physical Quantities Measurement, Hubei Key Laboratory of Gravitation and Quantum Physics, School of Physics, Huazhong University of Science and Technology, 1037 Luoyu Road, Wuhan 430074, People's Republic of China}%
\date{\today}

\begin{abstract}

We report on the development and phase noise performance of a 9.1926 GHz microwave frequency synthesizer to be used as the local oscillator for a Cs fountain clock. It is based on frequency multiplication and synthesis from an ultralow phase noise 5 MHz Oven Controlled Crystal Oscillator (OCXO) and 100 MHz Voltage Controlled Crystal Oscillator (VCXO).The key component of the frequency multiplication is a non-linear transmission-line (NLTL) used as a frequency comb generator. The phase noise of the synthesizer is improved by carefully optimizing the input power, the input and output impedances of the NLTL. The absolute phase noises of the 9.1926 GHz output signal are measured to be $-64$ dBc/Hz, $-83$ dBc/Hz, $-92$ dBc/Hz, $-117$ dBc/Hz and $-119$ dBc/Hz at 1 Hz, 10Hz, 100Hz, 1 kHz and 10 kHz offset frequencies, respectively. The residual phase noise of the synthesizer is measured to be $-82$ dBc/Hz at 1 Hz offset frequency. The measurement result shows that the absolute phase noise at the frequency range of 1 - 100 Hz is mainly limited by the phase noise of the OCXO. The contribution of the absolute phase noise to the fountain clock short-term frequency stability is calculated to be $7.0 \times 10^{-14} \tau^{-1/2}$. The residual frequency stability of the synthesizer is measured to be $1.5 \times 10^{-14} \tau^{-1/2}$, which is consistent with the calculated frequency stability due to the residual phase noise of the synthesizer. Meanwhile we designed and realized an interferometric microwave switch in the synthesizer to eliminate the frequency shifts induced by the microwave leakage. The extinction ratio of the switch is measured to be more than 50 dB. In the scheme, we use only commercially available components to build the microwave frequency synthesizer with excellent phase noise performance for high-performance Cs fountain clocks.
\end{abstract}

\maketitle

\section{INTRODUCTION}

The Second was defined as `the duration of 9 192 631 770 periods of the radiation corresponding to the transition of the cesium 133 atom'\cite{Resolution11967}. Since then, Cs clocks have been used as primary frequency standards to realize the definition of the Second. Along with the appearance of the laser cooling technique\cite{Bjorkholm1987}, the laser cooled neutral atoms have attracted the attention of the time and frequency community for its possible application to the development of a new generation of Cs primary frequency standards. Indeed, in 1995, Clarion's group in LNE-SYRTE developed the first laser-cooled Cs fountain clock and immediately demonstrated a better accuracy than thermal Cs beam standards\cite{Clairon1995}. In the following years, many national metrology institutes worldwide started to develop their own Cs fountain clocks. Up to now, data from 14 Cs fountain clocks have been reported to the International Bureau of Weights and Measures (BIPM) and more than one order of magnitude improvement in the accuracy has been realized in the definition of the SI Second\cite{Fang2015,Levi2014,Thomas2014,Szymaniec2010,Gerginov2010,Kumagai2008,Szymaniec2005,Jefferts2003,Weyers2001,Levi2006,Takamizawa2015}. Among all these Cs fountain clocks, the Cs fountain clock NIST-F2 developed by the National Institute of Standards and Technology (NIST)\cite{Thomas2014} has the best type B fractional frequency uncertainty of $1.1 \times 10^{-16}$. The Cs fountain clock developed by LNE-SYRTE\cite{Guena2012} based on a microwave frequency synthesizer from a cryogenic sapphire oscillator (CSO) has the best short-term frequency stability of $1.6 \times 10^{-14} \tau^{-1/2}$.

For Cs fountain clocks, the short-term frequency stability not only limit the type A fractional frequency uncertainty, but also limit the time for evaluating the type B fractional frequency uncertainty. Consequently, it is beneficial to improve the short-term frequency stability of the Cs fountain clocks. The Allan standard deviation of the fractional frequency fluctuations of an atomic fountain clock can be expressed as\cite{Santarelli1999}
\begin{equation}
 {\sigma _y}\left( \tau  \right) \!=\! \frac{1}{{\pi {Q_{at}}}}\sqrt {\frac{{{T_c}}}{\tau }} {\left( {\frac{1}{{{N_{at}}}} \!+\! \frac{1}{{{N_{at}}{n_{ph}}}} \!+\! \frac{{2\sigma _{\delta N}^2}}{{N_{at}^2}} \!+\! \gamma } \right)^{1/2}},
\end{equation}
where $\tau$ is the average time in seconds. \emph{T}$_c$ is the fountain cycle duration ($\sim$1.5 s). \emph{Q}$_a$$_t$=$\nu$$_0$/$\Delta\nu$ is the atomic quality factor. The first term in the brackets is due to the quantum projection noise\cite{Itano1993}. The middle two terms are due to the photon shot noise of the detected fluorescence and the detection system noise, respectively. They are generally less than that of the quantum projection noise. The last term $\gamma$ is due to the frequency noise of the interrogation oscillator through the Dick effect\cite{Dick1987}. The frequency stability limitation of an atomic clock due to the Dick effect is calculated as\cite{Rovera1996,Santarelli1998}
\begin{small}
\begin{equation}
 \sigma _y^{Dick}\left( \tau  \right) \!= \!{\left( {\frac{1}{\tau }\sum\limits_{m\! =\! 1}^\infty  {\frac{{g_m^2}}{{g_0^2}}{S_y}\left( {m/{T_c}} \right)} } \right)^{1/2}}{\rm{ \!= \! }}{\left( {\frac{1}{\tau }\sum\limits_{m \!=\! 1}^\infty  {\frac{{g_m^2}}{{g_0^2}}{S_y}\left( {m{f_c}} \right)} } \right)^{1/2}},
\end{equation}
\end{small}
where the parameters \emph{g}$_m$ and \emph{g}$_0$ are defined from the sensitivity function \emph{g}(t). Here \emph{S}$_y$(\emph{f}$_c$)=(\emph{f}/\emph{f}$_0$)S$_\varphi$(\emph{f}$_c$) is the power spectral density (PSD) of microwave oscillator fractional frequency fluctuations at the offset frequency \emph{f}$_c$. \emph{S}$_\varphi$(\emph{f}$_c$) is the PSD of phase fluctuations. Since \emph{f}$_c$=1/\emph{T}$_c$ $\sim$ 0.67 Hz, the contribution to the short-term frequency stability of the Cs fountain due to the phase noise at smaller offset frequency of the microwave frequency synthesizer is larger than that from larger offset frequency. The microwave frequency synthesizer based on an ultrastable laser\cite{Lipphardt2016,Millo2009,Fortier2011,Weyers2009,Millo2009,Lipphardt2008} or a cryogenic sapphire oscillator (CSO)\cite{Mill2009,Abgrall2016,Takamizawa2014,Watabe2005,Chambon2007,Heo2016} has superior phase noise and frequency stability properties. Thus, the Dick effect contribution can be considerably reduced and reach the quantum projection noise limit using the ultrastable and ultralow phase noise microwave frequency synthesizers from the schemes. Nevertheless, these systems remain complex, expensive and bulky. Therefore, the generation of microwave signals referenced to an ultralow noise quartz crystal oscillator remains a well-adapted solution\cite{Heavner2005,Gupta2007,Francois2014,Francois2015,Boudot2009}.

This paper describes the development of a 9.1926 GHz microwave frequency synthesizer for high-performance Cs fountain clock. The generation of the microwave signal is based on frequency multiplication and synthesis from an ultralow phase noise 5 MHz OCXO and 100 MHz VCXO. The key component of frequency multiplication is a non-linear transmission line (NLTL) used as a frequency comb generator that was also used in Ref. 33 and 35. The phase noise of the synthesizer is efficiently improved by carefully optimizing the input power, the input and output impedances of the NLTL. And in our scheme, we omitted some unnecessary filters and amplifiers which could introduce additional phase noise and did not improve phase noise performance of the synthesizer. And almost all components are commercially available. The different components are easily connected by semi-rigid cables with SMA adaptors whose shielding factor is more than 100 dB to avoid microwave leakage and signal crosstalk. Based on the design, we realized a excellent phase noise and residual frequency stability performances microwave frequency synthesizer of 9.1926 GHz for application to high-performance Cs fountain clocks. The contribution to short-term frequency stability due to Dick effect is estimated to be better than $7.0 \times 10^{-14} \tau^{-1/2}$. And the residual frequency stability of the synthesizer is measured to be $1.5 \times 10^{-14} \tau^{-1/2}$, which is consistent with the calculated stability due to the residual phase noise of the synthesizer. Meanwhile, by using an interferometric switch, an extinction ratio of more than 50 dB is realized to suppress the microwave leakage frequency shifts of the Cs fountain clocks.

\section{ARCHITECTURE OF THE MICROWAVE FREQUENCY SYNTHESIZER}
Figure. 1 shows a detailed scheme of the microwave frequency synthesizer. The whole setup can be divided into six main blocks. The key element of the synthesizer is a 5 MHz OCXO (Microsemi 1000C) with ultralow phase noise at the 1-100 Hz offset frequencies. The output power of the OCXO is about $+14$ dBm and its tuning voltage control frequency sensitivity was measured to be about $0.25$ Hz/V. In order to suppress the additional noise contribution of the phase-frequency detector\cite{Brennan2001}, The output 5 MHz signal of the OCXO is first frequency-doubled to 10 MHz using an active frequency-doubler (SDI FS020-5) and then further increased to 100 MHz by an active frequency-multiplier (SDI FS100-10). The output power of 100 MHz is about $+12$ dBm. The 100 MHz signal is then split into two arms. In the first arm, the 100 MHz is phase locked to an active Hydrogen maser (Microsemi MHM2010) by an analog phase locked loop (APLL). The locked loop bandwidth is set at about 0.1 Hz (lock time $\sim$ 10 s) to maintain the excellent phase noise of the OCXO and good long-term frequency stability of Hydrogen maser at the same time. The second arm 100 MHz is used to phase lock a 100 MHz VCXO whose phase noise is excellent at offset frequency range of 100 Hz - 200 kHz. The output 100 MHz signal of locked-VCXO is then frequency doubled to 200 MHz using a passive frequency-doubler (Mini-circuits ZX90-2-13+). The 200 MHz signal is divided into three arms utilizing a 3-way power splitter (Mini-circuits ZCSC-3-R3+). The first arm 200 MHz signal is amplified with two stage low noise amplifiers (Mini-circuits ZX60-P105LN+ and ZFL-1000VH2+) and then attenuated with fixed attenuators to a power of $21.5$ dBm. The 200 MHz signal is used to drive a NLTL comb generator device (MACOM MLPNC-7100) that generates harmonics up to 20 GHz. The output 9.0 GHz harmonic signal is bandpass filtered (BPF) with a 100 MHz cavity-filter and then amplified to a power of about 0 dBm with two ultra-low phase noise microwave amplifiers (Hitter HMC-C072).

The NLTL component was found to be very critical\cite{Francois2014}. The input power, input and output impedances matching of the NLTL can severely affect the phase noise of the output signal. Thus, the optimal input power of 200 MHz signal is determined by measuring the phase noise of the output 9.0 GHz signal. The input impedance of the NLTL is matched with an L-network circuit. The microwave isolators and fixed attenuators are added at the output of the NLTL to help for output impedance matching. Microwave isolators are also used to prevent the undesired power feedback.

Once the best configuration was found, we observed an excellent performance of phase noise. The output 9.0 GHz signal of the NLTL is mixed with the 8.9926 GHz signal coming from a dielectric resonator oscillator (MD DRO-1010-08.992), to produce a frequency beatnote at $7.368$ MHz. The $7.368$ MHz signal is directly compared with the 7.368 MHz signal from a direct digital synthesizer (DDS AD9852). The DDS is referenced to the second arm 200 MHz signal. So the DRO is phase-locked to the locked-100 MHz signal. Here we omitted the low-pass filter and amplifier for 7.368 MHz because the frequency of another mixed harmonic signal is much higher than the pull-in bandwidth of the PLL. The 7.368 MHz sine wave signals from DDS and the beatnote 7.368 MHz signal are all converted to square wave signals with two differential receivers whose output powers do not change with variations of input powers. The two square wave signals are phase detected by a Phase-Frequency Detector (PFD MCK12140). The output 8.9926 GHz of locked-DRO is mixed with the third 200 MHz signal which is amplified and then went through an interferometric switch to produce the 9.1926 GHz output signal. This signal is used as the local oscillator with low phase noise performance for Cs fountain clocks.

The interferometric switch is utilized to suppress the microwave leakage frequency shift. Concerns about phase transients have been addressed before\cite{Santarelli2009,Yuzhang2014}. In our scheme, we designed a interferometric switch as shown in Fig. 2. The 200 MHz signal is split into two branches with a 2 way-$90^\circ $ power splitter (Mini-circuits JSPQ-350+). The $90^ \circ $-branch is phase shifted by a voltage-controlled phase shifter (Mini-circuits SPHSA-251+) and then attenuated by a voltage variable attenuator (Mini-circuits HVA-451+), and finally through a TTL- driven switch with a switching time of 7 ns, the signal is power combined with the $0^ \circ $-branch 200 MHz signal with a power combiner (Mini-circuits SYPS-2-52HP+). The switch is on when the TTL level is high and off when the TTL level is low. The extinction ratio of the 9.1926 GHz microwave signal is measured to be larger than 50 dB by adjusting the power and phase of the two branches input signals of the power combiner.

\begin{figure}
\includegraphics[width=9cm]{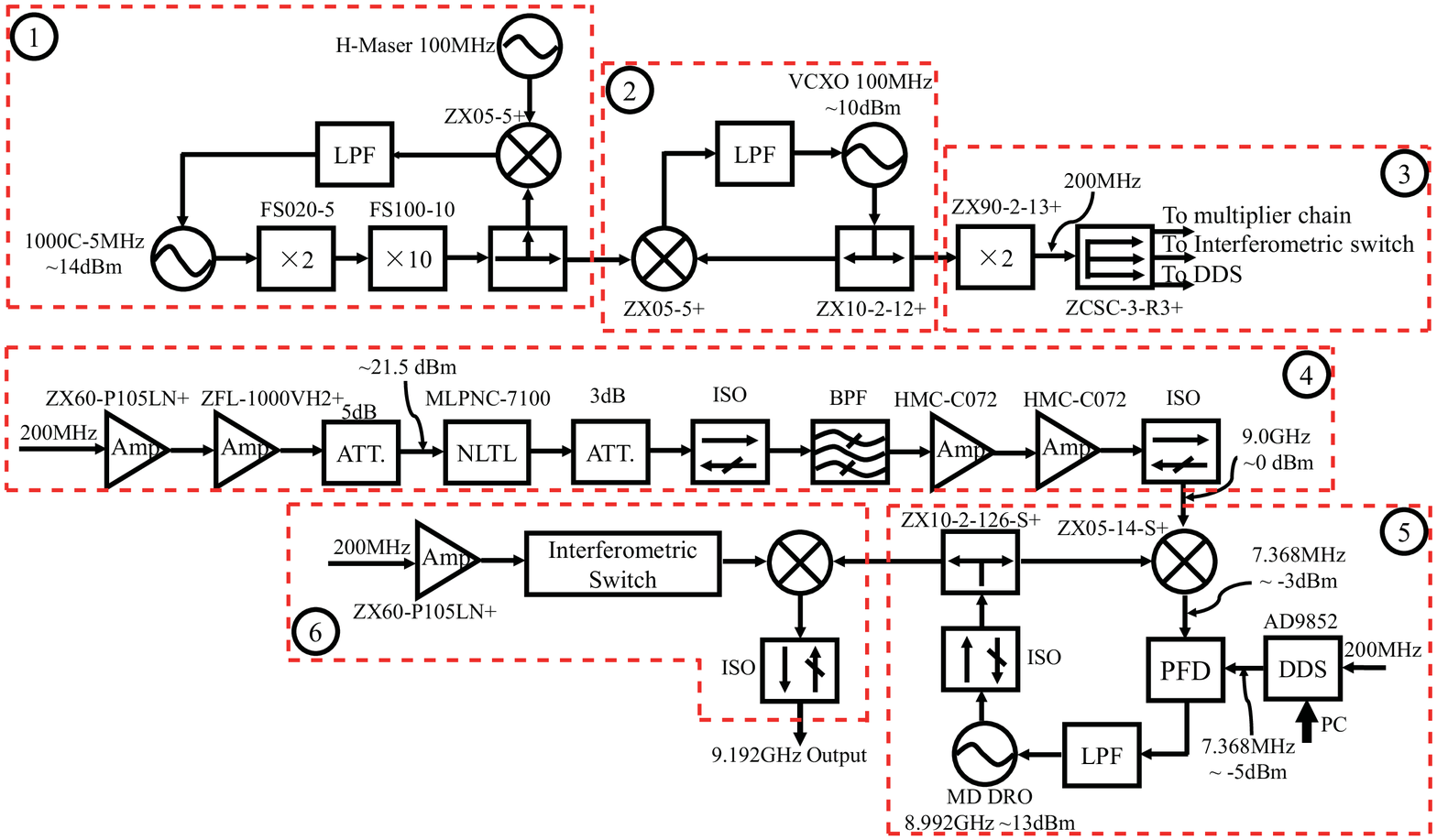}
\caption{(Colors online) Architecture of the microwave frequency synthesizer. Six main blocks are indicated. 1: 5 MHz OCXO phase locked to H-maser; 2: 100 MHz VCXO phase locked to OCXO; 3: 100-200 MHz frequency doubling stage; 4: NLTL-base 200 MHz- 9.0GHz chain; 5: 8.992 GHz DRO phase lock loop; 6: switch controlled 9.1926 GHz output signal.}
\end{figure}

\begin{figure}
\includegraphics[width=6cm]{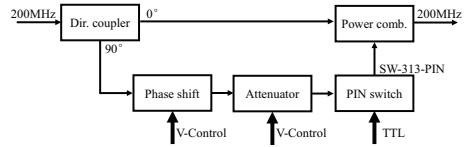}
\caption{(Colors online) Setup of the 200MHz interferometric switch.}
\end{figure}

\section{PERFORMANCE OF THE MICROWAVE FREQUENCY SYNTHESIZER}
Figure 3 shows an interior photograph of the developed microwave frequency synthesizer. All components except 5MHz OCXO are fixed in a 3U-case. The performances of the microwave frequency synthesizer have been evaluated, such as RF spectrum, phase noise, Dick effect and residual frequency stability. Presented here are the detailed measurements of the performances of the microwave frequency synthesizer. The frequency microwave synthesizer will be used in our Cs fountain clock development.

\begin{figure}
\includegraphics[width=6cm]{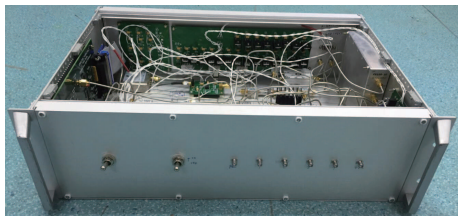}
\caption{(Colors online) A photograph of the microwave frequency synthesizer.}
\end{figure}

\subsection{Spectrum of the microwave signal and evaluation of related shifts}
The microwave related shifts induced by microwave power, microwave leakage, microwave phase variations and spurious components of the RF spectrum are some of the important uncertainty terms for Cs fountain clocks\cite{Fang2015,Levi2014,Thomas2014,Gerginov2010,Li2011}. The microwave frequency synthesizer was primarily designed to suppress these frequency shifts. The measured RF spectrum is shown in Fig. 4. The output power of the 9.1926 GHz signal is about -20 dBm, and there are no spurious components.  From the theoretical analysis, the related shifts induced by spurious and microwave leakage can be neglected. The microwave power and phase variations shifts will be estimated in future experiments.
\begin{figure}
\includegraphics[width=9cm]{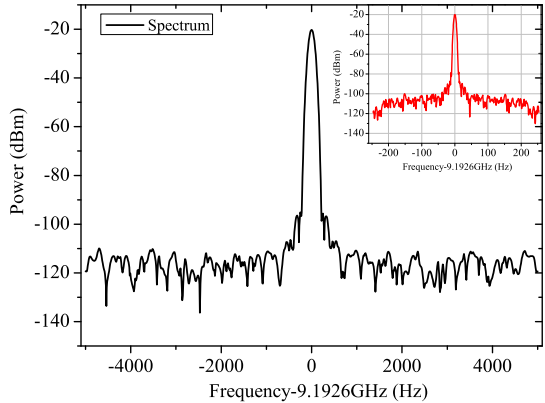}
\caption{(Colors online) RF spectrum of the microwave signal.}
\end{figure}

\subsection{Phase noise performances measurements}

To choose the optimal parameters of the PLL, we first measured the absolute phase noise performance of the signal sources of 5 MHz OCXO, 100 MHz VCXO and 8.9926 GHz DRO. The measurements are performed with a phase noise measurement system (Agilent E5500A) using the `phase detector with reference source' technique. Two similar signal sources are sent to the E5500A. The one is used as the device under test (DUT), and the other is used as the reference. The measured result is shown in Fig. 5. The carrier frequencies are all normalized to 9.0 GHz to analyze and compare results directly. The figure shows that the phase noise of the 5 MHz OCXO is ultralow at offset frequencies lower than 100 Hz. There is very good phase noise performance for the 100 MHz VCXO at offset frequencies between 100 Hz and 200 kHz. And the phase noise of 8.9926 GHz DRO is excellent at offset frequencies beyond 200 kHz. Thus, we designed the bandwidth of the PLL to be about 100Hz and 200 kHz for the 100 MHz VCXO and the DRO respectively to obtain the best phase noise performance.

\begin{figure}
\includegraphics[width=9cm]{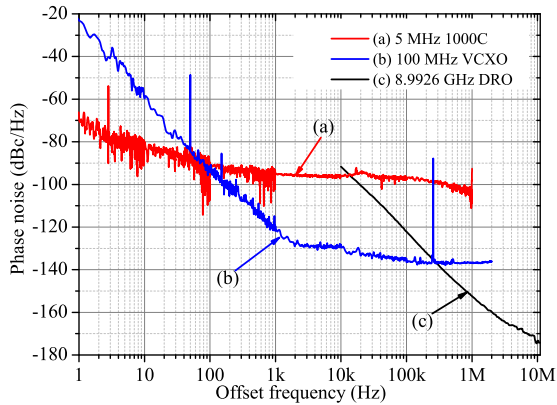}
\caption{(Colors online) Absolute phase noise of the 5 MHz OCXO, the 100 MHz VCXO and the 8.9926 GHz DRO, The carrier frequencies are all normalized to 9.0 GHz.}
\end{figure}
In the 200 MHz - 9.0 GHz frequency multiplication chain, we use an ultralow phase noise NLTL comb generator MLPNC 7100 with a flexible range of input frequency of 100 - 400 MHz and input power of 18 - 24 dBm. We measured the input impedance at 200 MHz of the NLTL using a network analyzer (Agilent 5061B). The complex impedance was measured to be \emph{Z}$_N$$_L$$_T$$_L$=7+j35 $\Omega$, which is not compatible with the required 50 $\Omega$. The magnitude of the reflection coefficient \emph{S}$_1$$_1$ is only -4 dB. It means that almost 60\% of the 200 MHz output power is reflected back and cannot be input to the NLTL. Thus, we designed a simple impedance matching L-circuit (IMC) to match the NLTL input impedance to the 200 MHz output. The measured \emph{S}$_1$$_1$ parameter with and without the IMC is shown in Fig. 6. As shown in Fig. 6, the magnitude of the \emph{S}$_1$$_1$ was improved to be about -35 dB by carefully adjusting the C1 and C2. At the same time, we measured the absolute phase noise of the output 9.0 GHz of the NLTL using a Phase Noise Analyzer (R$\&$S FSWP26 with option B60) using cross-correlation techniques for enhancing the sensitivity. It is found that the phase noise can be improved about 3 - 5 dB in the 200 Hz - 200 kHz offset frequency range because of the IMC as shown in Fig. 7.

For further investigating the phase noise performance of the NLTL, figure 8 reports the absolute phase noise of the output 9.0 GHz signal of the NLTL for different input powers. In this experiment, it is found that the phase noise of the output signal of the NLTL is severely affected by the input power at the offset frequency range of 200 Hz - 200 kHz, and there is an optimal input power to achieve the best phase noise performance for the NLTL. In our measurement, the best phase noise performance is obtained for an input power of about 21.5 dBm.

\begin{figure}
\includegraphics[width=9cm]{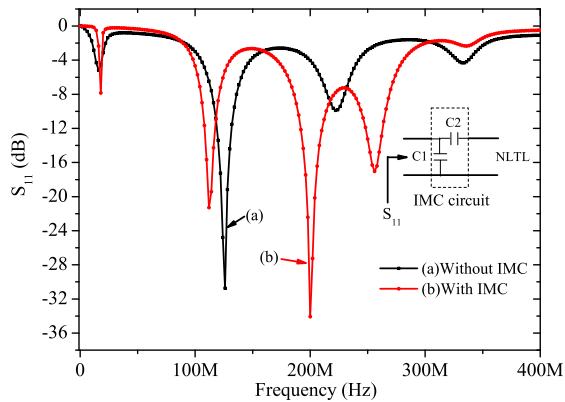}
\caption{(Colors online) The S$_1$$_1$ parameters of NLTL input terminal with and without IMC.}
\end{figure}
\begin{figure}
\includegraphics[width=9cm]{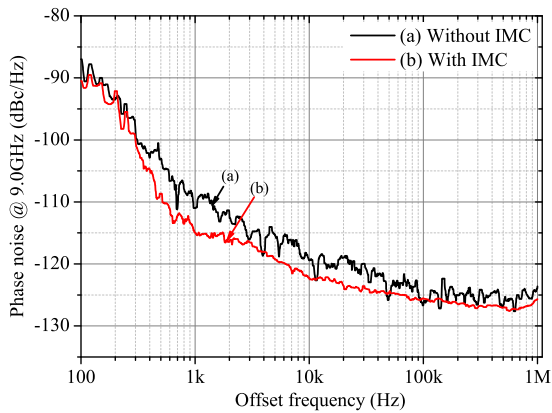}
\caption{(Colors online) The absolute phase noise of the output 9.0 GHz signal of the NLTL with and without IMC.}
\end{figure}
\begin{figure}
\includegraphics[width=9cm]{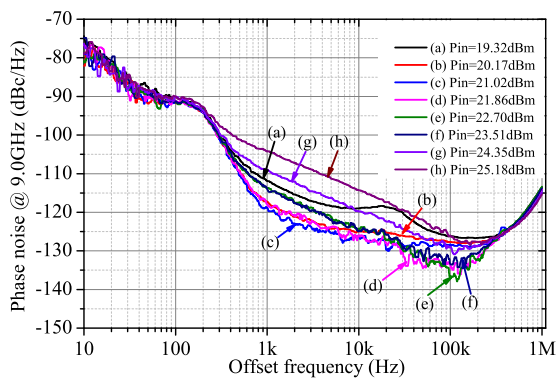}
\caption{(Colors online) Absolute phase noise at 9.0 GHz of the NLTL output versus the NLTL input 200 MHz signal power.}
\end{figure}
After choosing the optimal input parameters of the NLTL, the phase noise of the 9.0 GHz signal still deteriorate after passing the BPF and the microwave amplifier. However, the residual phase noises of the BPF and the microwave amplifier are measured to be very low at the carrier frequency of 9.0 GHz as shown in Fig. 9. We note that the output impedance of the NLTL is also not 50 $\Omega$ and cannot match with the input impedance of the BPF that is why the phase noise is worsen. A 1 dB fixed attenuator is helpful for impedance matching at the microwave frequency range. At first, we measured the absolute phase noise of the 9.0 GHz signal behind microwave amplifier by adding different amounts of fixed attenuators between the NLTL and the BPF. Figure 10 shows that the phase noise of the 9.0 GHz signal is significantly improved after at least three 1 dB fixed attenuators are added. And then we measured the residual phase noise with regard to the optimal parameters of the NLTL as shown in Fig. 10. As we can see, the absolute phase noise nearly reaches the residual phase noise of the frequency multiplication chain beyond 4 kHz offset frequency. But it is slightly smaller than the residual phase noise at the frequency range of 600 Hz - 4 kHz. The reason is that the absolute phase noise is measured at the locked condition. To certain extents, the absolute phase noise is suppressed under locked condition.

\begin{figure}
\includegraphics[width=9cm]{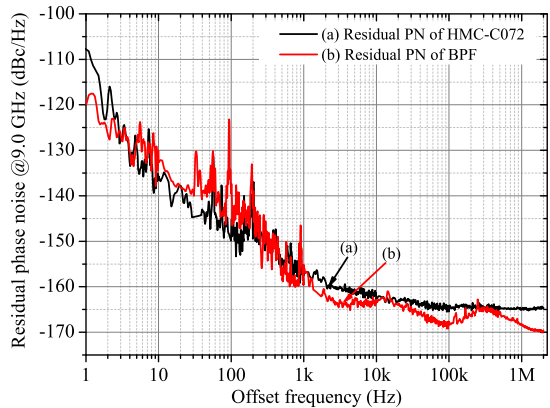}
\caption{(Colors online) Residual phase noise of the bandpass filter and microwave amplifier.}
\end{figure}
\begin{figure}
\includegraphics[width=9cm]{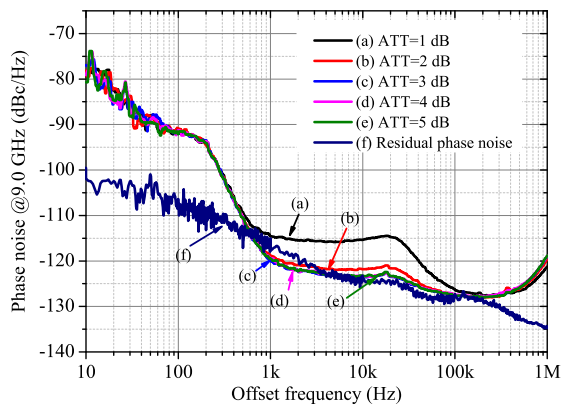}
\caption{(Colors online) Absolute phase noise of the 9.0 GHz behind the microwave amplifier versus number of fixed attenuator inserted between the NLTL and the BPF.And the residual phase noise of the 200MHz to 9.0 GHz frequency multiplication chain.}
\end{figure}
Figure 11 shows the absolute phase noise performances of the key signals of the microwave frequency synthesizer at 7.368 MHz, 100 MHz, 200 MHz, and 9.1926 GHz, respectively. The phase noise of the output 100 MHz signal of the locked VCXO is excellent without introducing additional noise other than near the frequency of locking bandwidth ($\sim 100$ Hz). For offset frequencies smaller than 1 kHz, as expected, the phase noise of the 200 MHz output signal is measured to be 6 dB higher than the phase noise of the 100 MHz signal. However, for offset frequencies larger than 1 kHz, the phase noise of the amplified 200 MHz output signal is degraded additional 6 - 10 dB which is introduced by the ZX60-P105LN+ and ZFL-1000VH2+ amplifiers. The phase noise of the output 9.1926 GHz signal does not achieve the theoretical noise level that is 20logN degraded from the 200 MHz signal at offset frequencies beyond 1 kHz. The excess phase noise may be still induced by the imperfect matched impedances between the NLTL and the BPF. The residual phase noise of the 9.1926 GHz signal is measured to be -82 dBc/Hz at 1 Hz offset frequency. And the measurement result shows that the absolute phase noise of the synthesizer at the frequency range of 1 - 100 Hz is mainly limited by the phase noise of the OCXO. The DDS phase noise has a bump around offset frequency range of 10 - 100 kHz. It is found that is induced by a 200 MHz band pass filter (Mini-circuits BPF-199+). Fortunately, the phase noise of DDS realized by us is not the limitating factor for the output 9.1926 GHz\cite{Du2017}.

\begin{figure}
\includegraphics[width=9cm]{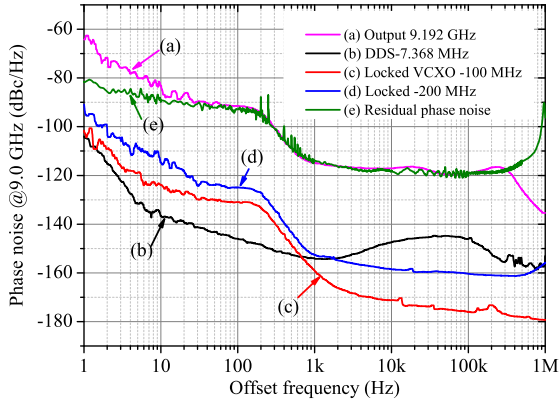}
\caption{(Colors online) Absolute phase noise performances of the key signals of the microwave frequency synthesizer. (a) DDS at 7.368 MHz, (b) locked VCXO at 100 MHz, (c) 200 MHz signal behind the amplifier, (d) output 9.1926 GHz signal of the synthesizer, (e)residual phase noise of the synthesizer.}
\end{figure}
\subsection{Dick effect contribution and residual frequency stability}

For typical operation parameters of a Cs fountain clock of Ramsey linewidth $\Delta \nu $=1 Hz, operation cycle time \emph{T}$_c$=1.5 s, microwave interrogation time $\tau$=14 ms, and microwave power \emph{b}$\tau$=$\pi$/2, it is calculated that frequency stability is limited at $7.0 \times 10^{-14} \tau^{-1/2}$ due to the absolute phase noise of the microwave frequency synthesizer through Dick effect\cite{Santarelli1998}. We measured the residual frequency stability of the microwave frequency synthesizer. The experiment setup is shown in Fig. 12. A common 5 MHz signal from the source OCXO is split into three arms. Two arms are sent into two identical synthesizers. The output signal frequencies at 9.192 GHz and 9.193 GHz are mixed to generate a beatnote at 1 MHz that is low-pass filtered and counted with a Microsemi 5120A phase noise and Allan deviation test set. The third arm 5 MHz signal is used as the reference signal of the 5120A.

Figure 13 reports the measured residual frequency stability of the beatnote and the computed Dick effect limitation. The residual frequency stability is measured to be $7.8 \times 10^{-15}$ at 1 s. The residual frequency stability limitation of the synthesizer is at the level of $1.5 \times 10^{-14} \tau^{-1/2}$ that is consistent with the theoretical calculation due to the residual phase noise of the synthesizer. But there is a big bump between average times of 100 s to 10000 s. Further investigation shows that the bump is mainly caused by the two active frequency multipliers FS020-5 and FS100-10. At the same time, we monitored the environment temperature and calculated its stability. The results show that the bump did not correlate with the environment temperature fluctuation. In our application, the measured residual frequency stability is still below the Dick effect contribution.

\begin{figure}
\includegraphics[width=6cm]{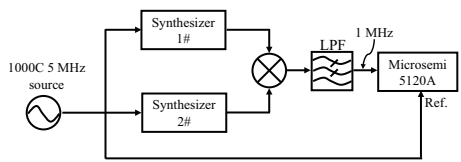}
\caption{(Colors online) Experiment setup used for measurement of the microwave frequency synthesizer residual frequency stability.}
\end{figure}
\begin{figure}
\includegraphics[width=9cm]{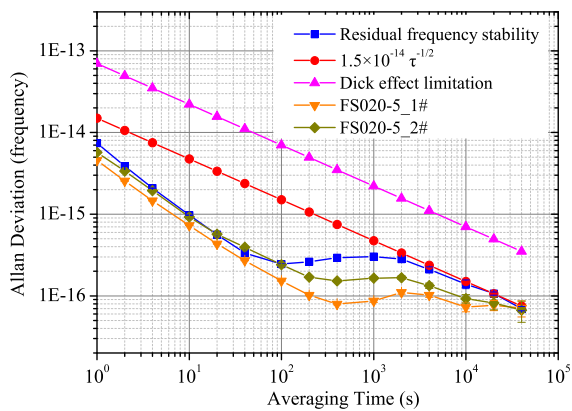}
\caption{(Colors online) Residual frequency stability and theoretical dick effect limitation of the microwave frequency synthesizer.And the residual frequency stability of the FS020-5 and environment temperature.}
\end{figure}

\section{CONCLUSION}

In conclusion, we reported the development and phase noise performances of a 9.1926 GHz microwave frequency synthesizer to be used as the local oscillator in a high-performance Cs fountain clock. It is based on frequency multiplication and synthesis from an ultralow phase noise 5 MHz OCXO and 100 MHz VCXO using a NLTL-based chain.The phase noise of the synthesizer is efficiently improved by carefully optimizing the parameters of the NLTL. Absolute phase noise performances of 9.1926 GHz output signal are $-64$ dBc/Hz, $-83$ dBc/Hz, $-92$ dBc/Hz, $-117$ dBc/Hz and $-119$ dBc/Hz at 1Hz, 10Hz, 100Hz, 1 kHz and 10 kHz offset frequencies, respectively. The residual phase noise of the synthesizer is measured to reach $-82$ dBc/Hz at 1 Hz. And the measurement result shows that the absolute phase noise at the frequency range of 1 - 100 Hz is mainly limited by the phase noise of the OCXO. The short-term frequency stability is limited at $7.0 \times 10^{-14} \tau^{-1/2}$ due to the absolute phase noise of the synthesizer through Dick effect. And the residual frequency stability of the microwave frequency synthesizer is measured to be $1.5 \times 10^{-14} \tau^{-1/2}$. Meanwhile, we realized an interferometric switch of extinction ratio of better than 50 dB to eliminate the microwave leakage induced frequency shifts.

\section{ACKNOWLEDGMENTS}

This work was partially funded by the National Key R$\&$D Program of China (Grant No. 2017YFA0304400),  the National Natural Science Foundation of China (Grant No. 91336213), and the Post-doctoral Science Foundation of China (Grant No. 2017M612431).

\end{document}